\documentclass[12pt]{article}

\newcommand{\bbr}{I\!\! R}

\newcommand{\3}{$^3$}

\newcommand{\x}{arXiv:}
\newcommand{\m}{\mathrm}

\usepackage{graphicx}

\begin{document}
\thispagestyle{empty}
\begin{center}

\null \vskip-1truecm \vskip2truecm {\Large{\bf

The Phantom Divide in String Gas Cosmology

}}

\vskip1truecm {\large Brett McInnes} \vskip1truecm

 National University of Singapore

email: matmcinn@nus.edu.sg\\

\end{center}
\vskip1truecm \centerline{ABSTRACT} \baselineskip=15pt
\medskip
One of the main virtues of string gas cosmology is that it
resolves cosmological singularities. Since the Universe can be
approximated by a locally asymptotically de Sitter spacetime by
the end of the inflationary era, a singularity theorem implies
that these cosmologies effectively violate the Null Energy
Condition [not just the Strong Energy Condition]. We stress that
this is an extremely robust result, which does \emph{not} depend
on assuming that the spatial sections remain precisely flat in the
early Universe. This means, however, that it must be possible for
string cosmologies to cross the recently much-discussed
\emph{phantom divide} [from w $<$ $-$1 to w $>$ $-$1, where w is
the equation-of-state parameter]. This naturally raises the
question as to whether the phantom divide can be crossed again, to
account for recent observations suggesting that w $<$ $-$1 at the
present time. We argue that non-perturbative string effects rule
out this possibility, even if the NEC violation in question is
only ``effective".

 \vskip3.5truecm
\begin{center}

\end{center}

\newpage

\addtocounter{section}{1}
\section* {\large{1. String Gases, Topology, and the Null Energy Condition}}
The string gas cosmology of Brandenberger and Vafa
\cite{kn:brandvafa}\cite{kn:brandeasson}\cite{kn:brandkim} has
many attractions, but, among them, two stand out in particular: it
is claimed that this approach can explain the existence of three
large dimensions, and that it can abolish cosmological
singularities\footnote{See
\cite{kn:brandwat}\cite{kn:watson}\cite{kn:greene}\cite{kn:danos}\cite{kn:campos}
for a sample of recent relevant work, and
\cite{kn:brandnew}\cite{kn:brandnewer} for clear summaries, many
further references, and  reports on recent advances on various
problems connected with the theory.}. In this work we shall
investigate the precise meaning of this second statement, and
explore the consequences.

In more detail, string gas cosmology is an attempt to use
unconventional spacetime geometry --- the spatial sections have
the topology of a three-torus, in the simplest models --- to
obtain cosmological models in which the large dimensions are
distinguished from the small by means of the concrete physical
mechanism of string annihilation. Such cosmologies are generically
non-singular, because there are momentum modes with energies
quantised in multiples of 1/R, where R is the radius of a given
dimension in string units. In some sense which remains to be made
precise, any attempt to probe length scales below the self-dual
radius will turn out to be a probe of \emph{larger}, T-dual
scales. This is very desirable, but one would like to understand
the consequences; it is reasonable to suppose that there are costs
involved.

Because string gas cosmologies are fully non-singular, one must of
course expect that some condition of the classical cosmic
singularity theorem [\cite{kn:hawking}, page 266] is violated.
Since all of the other conditions are certainly satisfied
[genericity and, notably, the presence of \emph{compact} edgeless
achronal sets], the only possibility is that the Strong Energy
Condition [SEC] is violated. This must be connected with an early
period of inflation, which can in fact be achieved in string gas
cosmology \cite{kn:mazumdar1}\cite{kn:mazumdar2}. As is well
known, however \cite{kn:guth}, inflationary violations of the SEC
in themselves by no means guarantee the avoidance of
singularities; thus we need to investigate further.

A clue as to how to proceed is provided by the fact that, by the
end of the inflationary era, the local geometry of the Universe
closely resembles that of de Sitter spacetime; that is, from the
point of view of still earlier eras, spacetime is
\emph{asymptotically locally de Sitter} towards the future.
Combining this observation with the requirement that singularities
should not appear, we find that string gas cosmologies effectively
violate the \emph{Null Energy Condition} [NEC], not just the SEC.
This follows from the singularity theorem due to Andersson and
Galloway \cite{kn:andergall}. The most remarkable feature of the
theorem is that it allows us to reach this conclusion without
making any assumptions about the detailed geometry of the early
Universe. In fact, the crucial assumption is \emph{topological}:
it is that the spatial sections should have a topology \emph{of
precisely the kind assumed in string gas cosmology}. Thus, the
theorem is ideally suited to this kind of cosmology; and, since
the crucial condition is topological, the conclusion that the NEC
has to be effectively violated is very robust --- it does not
depend on an assumed FRW structure or on the details of the
Einstein equation.

In short, in string gas cosmology, a period of inflation has to be
preceded by a period of [effective] NEC violation.

Before we proceed, let us be clear about the meaning of
``effective" NEC violation. The Andersson-Galloway theorem
actually implies a purely geometric conclusion, the violation of
the \emph{Null Ricci Condition} [see below]. If the Einstein field
equations hold exactly, then a violation of the Null Ricci
Condition immediately implies a violation of the NEC, which is of
course a condition on the stress-energy-momentum tensor. In more
complex gravitational theories, the Einstein equations may not
hold exactly; but one can still write the equations in the
standard form, with an extra contribution to the ``right hand
side". In such cases a violation of the Null Ricci Condition can
\emph{appear} to cause a violation of the NEC, simply because [for
example] the standard Friedmann equations are assumed in
reconstructing the pressure of the dark energy from the spacetime
geometry. In other words, the geometry of the spacetime will be
such as would result from a violation of the NEC, \emph{even
though no true matter field actually violates that condition}.
This is what we mean by ``effective" NEC violation; it corresponds
to a violation of the Null Ricci Condition.

All violations of the NEC, true or effective, are harder to
account for theoretically than violations of the SEC
\cite{kn:NEC}. The study of NEC-violating cosmologies was
initiated in \cite{kn:caldwell}; they are often singular [see
\cite{kn:NOT}\cite{kn:wei2} for more recent references discussing
the singular case], but the fact that they \emph{can} be
non-singular and asymptotically locally de Sitter was pointed out
in \cite{kn:smash}, and this observation has been developed in
various ways in
\cite{kn:escape}\cite{kn:srivastava}\cite{kn:puxun}\cite{kn:odintsov}\cite{kn:wei}\cite{kn:quiros}.
``Phantom cosmologies" involving \emph{true} violations of the NEC
encounter well-known objections \cite{kn:carroll}, but these are
avoided by more sophisticated models in which the violation is
merely \emph{effective}\footnote{See \cite{kn:nojiri} for a
general perspective on effective violations of the NEC.}: for
example, in brane-world theories \cite{kn:varun}\cite{kn:varun2},
or through quantum effects \cite{kn:onemli1}\cite{kn:onemli2}, or
by considering string-motivated Gauss-Bonnet corrections
\cite{kn:tsujikawa} to the Einstein-Hilbert Lagrangian. String
theory \emph{apparently} permits true NEC violation in the very
early Universe; specifically, there are in fact explicit
NEC-violating cosmologies which have been constructed within
string theory: Kachru and McAllister \cite{kn:kachru} have given
such an example in the context of a warped deformed conifold in a
Calabi-Yau compactification of II B string theory\footnote{This is
an effective scalar-tensor theory. The very subtle ways in which
the NEC may or may not be violated in such theories is thoroughly
explored in \cite{kn:ENO}.}. However, whether string theory allows
NEC violations, even of the ``effective" type, when the full
non-perturbative theory is taken into account, remains an open
question.

Returning to the case of string gas cosmology: the era of NEC
violation we have been discussing must be very brief --- it must
come to an end and be replaced by an inflationary era
\cite{kn:mazumdar1} \cite{kn:mazumdar2}. Hence we should look for
[true or effective] NEC violating effects which are evanescent.
Notice too that if the inflationary era is governed by a dynamical
inflaton [as opposed to a pure cosmological constant] then it
follows that it must be possible for these cosmologies to make the
transition from w $<$ $-$1 to w $>$ $-$1, where w is the
equation-of-state parameter. [This question was first discussed in
the context of the ``phantom inflation" model
\cite{kn:piaozhou}\cite{kn:piaozhang}.] The ability of a
cosmological model to ``cross the phantom divide" in this way
 can be a very subtle matter, as
has been emphasised recently in the astrophysics literature
\cite{kn:sahni}\cite{kn:bofeng}\cite{kn:vikman}\cite{kn:hu}\cite{kn:doran}\cite{kn:cai}\cite{kn:zhang}.
[In fact, these works are concerned with a transition in the
opposite direction, but the same issues arise here.] Thus string
gas cosmology must belong to this special class of models. This is
a highly non-trivial conclusion.

Next, in view of the fact that the Universe is again violating the
SEC at the present time \cite{kn:riess}, it is natural to ask
whether the same is true of the NEC: can the NEC, like the SEC, be
noticeably violated over a lengthy period, including the present?
Granted that string gas cosmologies are capable of crossing the
``phantom divide" in one direction, it seems natural to suppose
that they can cross in the other, from w $\geq$ $-$1 to w $<$
$-$1. In fact, as Hu has recently stressed \cite{kn:hu}, such a
crossing must occur if, as has been proposed \cite{kn:sahni}, the
observational data suggest that w is evolving rapidly. The
question is far from academic, since a \emph{recent} period of NEC
violation is entirely compatible with the observational data
\cite{kn:steinhardt}. Furthermore, some analyses of the data
\cite{kn:peri} suggest that they directly favour a recent crossing
of the phantom divide. [However, other analyses, for example
\cite{kn:choudpad}\cite{kn:pad}\cite{kn:tegmark}, suggest that the
data are still compatible with a simple cosmological constant; the
question remains open.]

Since we are now considering NEC violation over vast periods, it
is clear that this second period of NEC violation must be of the
``effective" kind, as for example arises very naturally in
brane-world models \cite{kn:varun}\cite{kn:varun2}\footnote{True
NEC violation is also, however, possible on braneworlds: see
\cite{kn:calcagni} for a discussion.}. We nevertheless wish to
argue that non-perturbative string physics does \emph{not} allow
even \emph{effective} violations of the NEC for an extended
period, at least not in a way that could account for claimed
observations of w $<$ $-$1 at the present time. The relevant
mechanism is the ``Schwinger-like" production of brane-antibrane
pairs, in the manner first discussed in
\cite{kn:mich}\cite{kn:seiberg}\cite{kn:yau}\cite{kn:galloway},
and recently re-considered in detail by Maldacena and Maoz
\cite{kn:maoz} and others
\cite{kn:buchel}\cite{kn:mcinnes}\cite{kn:answering}\cite{kn:maoz3}\cite{kn:porrati}\cite{kn:reallyflat}.
The key property of this process that we need here is that it is
determined entirely by the spacetime geometry: it does not ``care"
whether the NEC violation is true or effective. We show that if
the NEC is violated in asymptotically locally de Sitter spacetimes
--- even if only effectively --- then when a brane-antibrane pair
is created, it is usually possible to move one member of the pair
into a region where the action becomes negative. This is a signal
of instability. The only way to avoid this instability is to
choose the parameters so that the NEC-violating geometry
approaches local de Sitter geometry extremely rapidly with the
expansion of the Universe. This is precisely what we need in the
earliest stages, to allow the transition to the inflationary era;
but such a rapid approach to a de Sitter-like state would render
this matter undetectable at the present time. \emph{In short,
string cosmology predicts that the dark energy satisfies w $\geq$
$-$1 at the present time, and that there has been no crossing of
the phantom divide since the very earliest era}.

We have based this discussion on string gas cosmology, but it is
relevant to any asymptotically locally de Sitter cosmology with
toral spatial sections; see for example
\cite{kn:paban}\cite{kn:elizalde}\cite{kn:durrer}. For discussions
of compact spatial sections in quantum cosmology, see
\cite{kn:zeldovich}\cite{kn:ruback}\cite{kn:martin}\cite{kn:coule},
and see \cite{kn:lindetypical} for a discussion of the
compatibility of compact spatial sections with inflation. Note
that, in all these applications, including string gas cosmology,
the spatial sections can be finite \emph{quotients} of tori; in
fact, some of these quotients are probably preferred to tori,
since the internal space in the phenomenologically most
interesting cases \cite{kn:easson}\cite{kn:easther} is a toral
orbifold rather than a torus. See \cite{kn:conway} for the
interesting details of these quotients and
\cite{kn:kehag}\cite{kn:reallyflat} for examples in which these
details are physically important. For the sake of simplicity,
however, we shall refer to all of these spaces as ``tori", though
``compact flat three-manifolds" would be more precise.

Before we proceed, we should clarify that neither this work nor
any of those cited depends on the question
\cite{kn:weeks}\cite{kn:luminet} as to whether a toral spatial
structure is \emph{directly observable} in, for example, the
cosmic microwave background. A priori, one does not expect this to
be possible, since the size of a flat torus is of course unrelated
to its curvature and can be freely prescribed independently of all
other parameters. An observation of toral structure would
therefore constitute yet another ``cosmic coincidence". In any
case, our discussion has no bearing on this question.

Finally, our viewpoint on the whole question of NEC violation will
be string-theoretic throughout. For other perspectives, see
\cite{kn:singh}.

\addtocounter{section}{1}
\section*{\large{2. The Global Formulation of String Gas Spacetimes}}
In this section, we give a more precise discussion of the
assertion that string gas cosmologies are \emph{non-singular}.
This will also permit us to state the Andersson-Galloway theorem.

A four-dimensional spacetime M$_4$ with Lorentzian metric
$g_{\mathrm{M}}$ is said \cite{kn:andergall} to have a {\em
regular future [past] spacelike conformal completion} if M$_4$ can
be regarded as the interior of a spacetime-with-boundary X$_4$,
with a [non-degenerate] metric $g_{\mathrm{X}}$ such that the
boundary is {\em spacelike} and lies to the future [past] of all
points in M$_4$, while $g_{\mathrm{X}}$ is conformal to $g_M$,
that is, $g_{\mathrm{X}}$ = $\Omega^2g_{\mathrm{M}}$, where
$\Omega$ = 0 along the boundary but d$\Omega \neq 0$ there. [There
is a straightforward modification which allows us to define what
it means for a spacetime to have a regular \emph{future and past}
spacelike conformal completion.]
\begin{figure}[!h]
\centering
\includegraphics[width=0.5\textwidth]{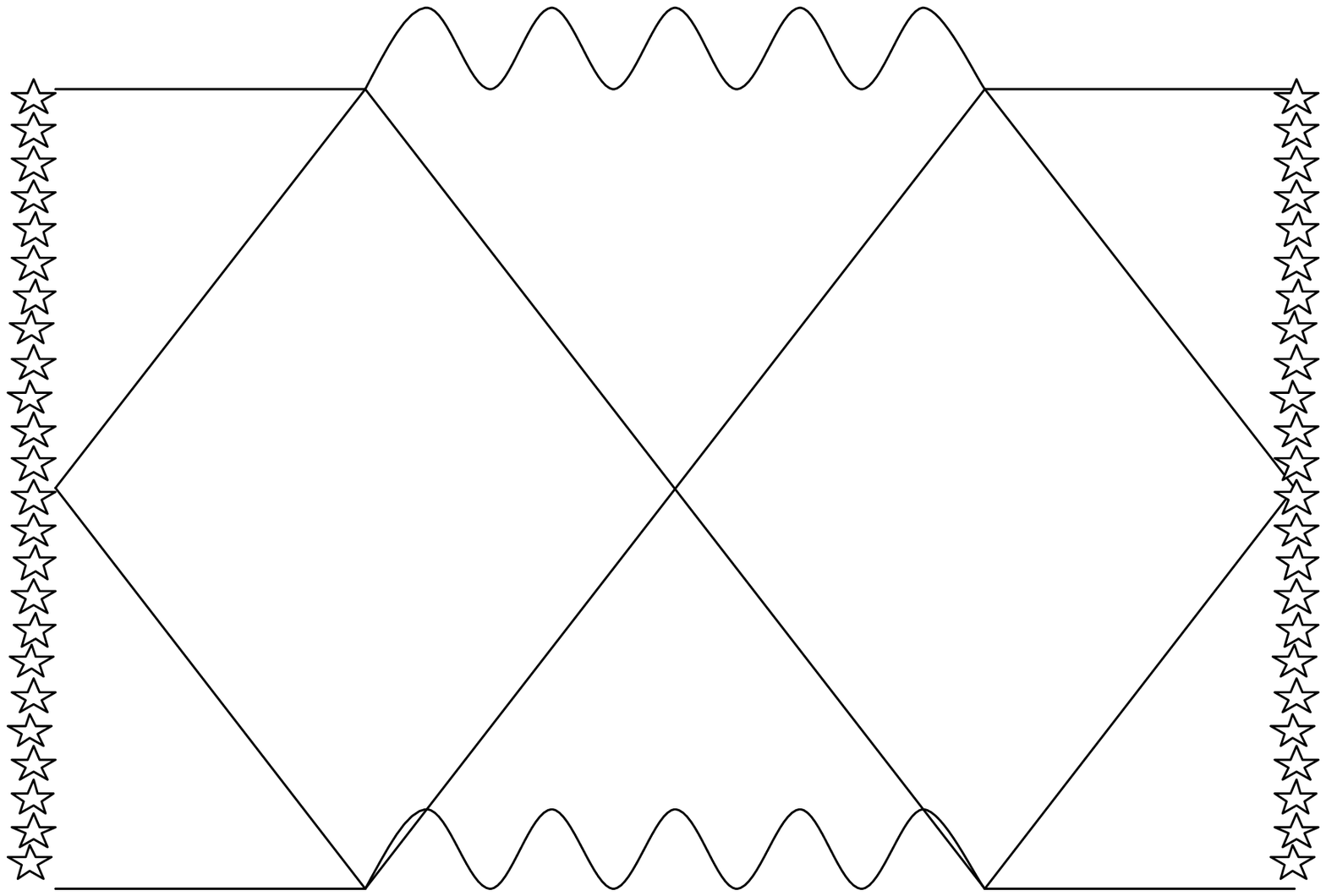}
\caption{Schwarzschild-de Sitter According to Schwarzschild and de
Sitter}
\end{figure}
This definition captures one's intuitive understanding of an
``asymptotically locally de Sitter" spacetime [towards the future,
the past, or both]. For it is possible to show that the curvature
of such a spacetime will increasingly resemble that of de Sitter
spacetime along a timelike geodesic that escapes to future or past
infinity. However, this is a strictly local statement; it is very
important for us to understand how different such a spacetime can
be from ordinary de Sitter spacetime.

An example will make the answer clear. Consider the Penrose
diagram given in Figure 1. This spacetime is constructed as
follows: first, recall that conventional de Sitter spacetime has
the topology $\bbr$ $\times$ S\3. We can however perform an
identification of antipodal points on the three-sphere, to obtain
a version with spatial sections isometric to the real projective
space $\bbr$P\3. [This is in fact the way that de Sitter spacetime
was interpreted by Schwarzschild and de Sitter --- see
\cite{kn:rp3} for a discussion of this, and \cite{kn:louko} for
further work in this direction.]

Now introduce an electrically neutral black hole into this version
of de Sitter spacetime. The result is as shown in the diagram: the
past and future singularities are shown as usual, together with
the expected tunnel between two copies of what would otherwise
have been the $\bbr$P\3 version of de Sitter spacetime. The stars
represent copies of the two-dimensional projective space
$\bbr$P$^2$; all other points represent S$^2$.

This spacetime has a regular future and past spacelike conformal
completion and so is asymptotically locally de Sitter by our
definition. That is, any observer who does not fall into the black
hole will eventually find that the geometry in his neighbourhood
is indistinguishable from that of de Sitter spacetime. The
\emph{global} structure, however, could hardly be more different
from that of de Sitter spacetime. Notice in particular that the
conformal boundary in this case is certainly not S\3 --- indeed,
it is not even compact, being four copies of an $\bbr$P\3 from
which a point has been deleted. [This non-compactness is closely
related to the fact that the spacetime is singular, since this
spacetime is globally hyperbolic
--- see \cite{kn:andergall}.] In fact, the definition of
``asymptotically locally de Sitter" allows for indefinite
acceleration in a wide variety of ways. The definition \emph{does}
however exclude spacetimes like the Nariai spacetime [see for
example \cite{kn:bousso}] which fail to induce a non-degenerate
conformal structure at infinity. This is justifiable on the
grounds that such spacetimes are non-generic
--- two directions in Nariai spacetime do not expand at all, and
hence they are ``squeezed out" at infinity \cite{kn:rp3}. The
definition also excludes spacetimes like the ones considered
recently by Freivogel and Susskind \cite{kn:freebird}, where
future infinity is not spacelike. These spacetimes accelerate but
not indefinitely.

In short, the global structure of an asymptotically locally de
Sitter spacetime can differ radically from that of de Sitter
spacetime. We intend to make use of this freedom to investigate
the global structure of string gas cosmologies.

A spacetime with a regular future [past] spacelike conformal
completion is said to be {\em future [past] asymptotically simple}
if every future [past] inextendible null geodesic has an endpoint
on future [past] conformal infinity. This formalizes the idea of a
non-singular asymptotically locally de Sitter spacetime --- no
photon is captured by a singularity in such a spacetime. The
spacetime pictured in Figure 1 is neither future nor past
asymptotically simple.

With these technicalities in hand, let us consider the global
structure of a simple string gas cosmology.

In string gas cosmologies, it does not make sense for the cosmic
scale factor to be smaller than a certain value: T-duality implies
that attempts to probe below this value will reveal physics at the
dual scale. In cosmology, this can be interpreted to mean that
attempts to probe before a certain \emph{time} will only produce
data corresponding to a later time. Thus, spacetime should simply
not exist for values of the scale factor corresponding to
distances below the self-dual radius. We can model this as
follows.

Let us begin with the more familiar case of de Sitter spacetime
itself. This is \emph{not} a string gas cosmology but its
simplicity will clarify the point we wish to make. The metric, in
terms of conformal time, together with the usual angular
coordinates on the three-sphere of radius L, is given by
\begin{eqnarray}\label{eq:AAA}
g(\mathrm{dS}_4) = {{\mathrm{L}^2} \over {\mathrm{sin}^2(\eta)}}[
\mathrm{d}\eta^2\; -\; \mathrm{d}\chi^2 \; -\;
\mathrm{sin}^2(\chi)\{\mathrm{d}\theta^2 \;+\;
\mathrm{sin}^2(\theta)\,\mathrm{d}\phi^2\}].
\end{eqnarray}
Here $\eta$ is defined in the open interval (0, $\pi$). Notice
that, in this signature, this is a metric of constant
\emph{negative} curvature\footnote{We are accustomed to thinking
of de Sitter spacetime as having \emph{positive} curvature, but
this is merely a question of convention, depending on the choice
of ($-\;+\;+\;+$) signature rather than ($+\;-\;-\;-$).}; the
cosmological constant takes the negative value $-$ 3/L$^2$.

Now perform an identification according to the isometry
\begin{equation}\label{eq:AAAA}
\Theta\;:\;\eta\;\;\rightarrow\;\;\pi\;-\;\eta\,.
\end{equation}
The result is an orbifold spacetime in which the universe begins
at the \emph{non-singular} spacelike surface at $\eta$ = $\pi$/2,
which is the distinguished surface corresponding to the smallest
possible size permitted in this geometry. The universe can never
be smaller than this, and there are no earlier events. This
suggests a simple way of explicitly implementing the
non-singularity of string gas cosmologies: the Universe should be
``created from nothing" at this surface.

Notice that the initial surface is not embedded arbitrarily in the
spacetime. The lower half of the Penrose diagram of ordinary de
Sitter spacetime is an exact copy of the upper half, so that the
surface $\eta$ = $\pi$/2 presents the same appearance from either
side. The extrinsic curvature of the surface $\eta$ = $\pi$/2 [in
the original spacetime, before the quotient is taken] is therefore
zero, since otherwise it would provide a way of distinguishing one
side from the other.
\begin{figure}[!h]
\centering
\includegraphics[width=0.7\textwidth]{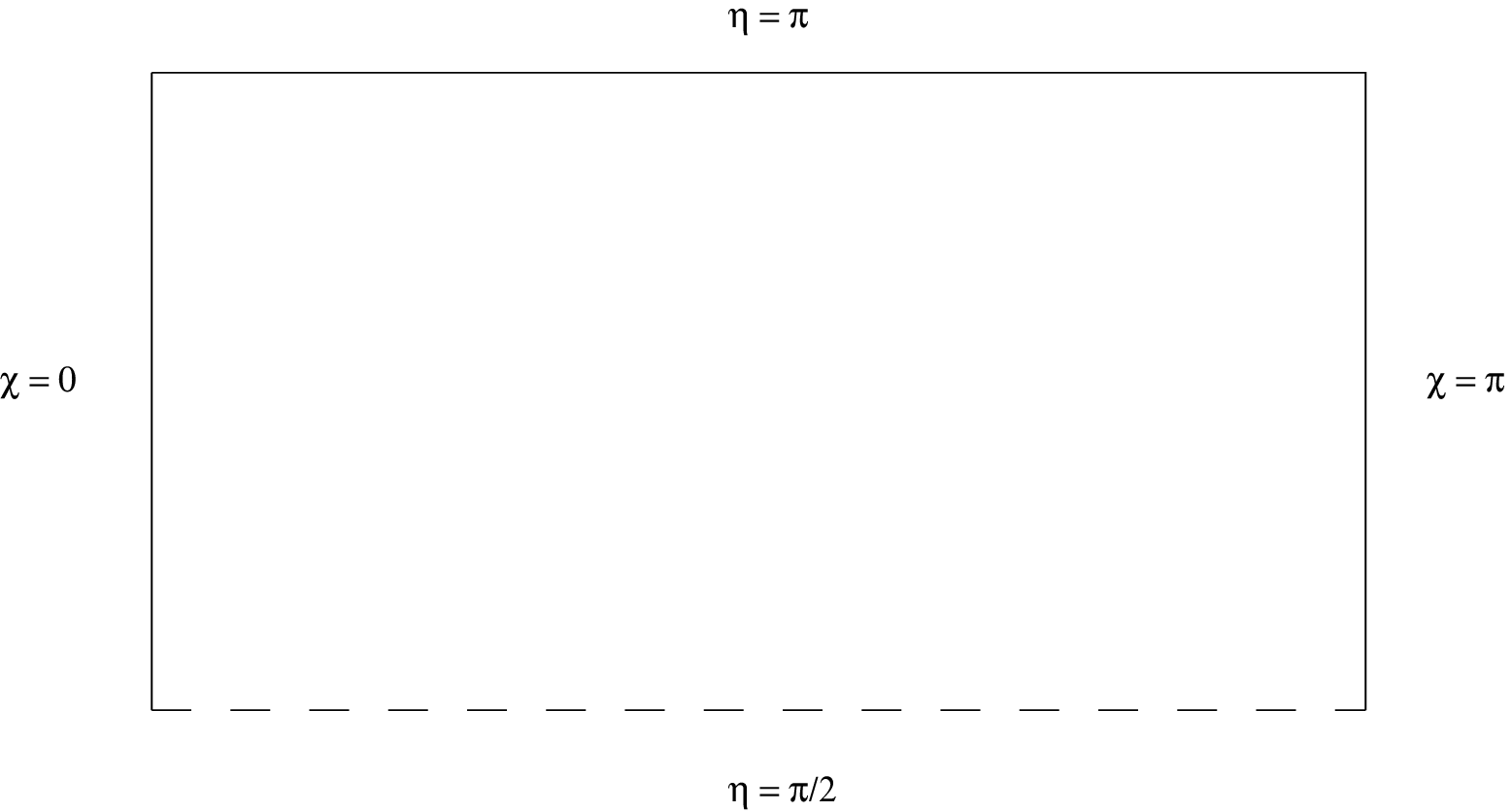}
\caption{Penrose diagram of the de Sitter temporal orbifold.}
\end{figure}
The Penrose diagram is as shown in Figure 2 [in the case where the
spatial topology is that of S$^3$ rather than $\bbr$P$^3$ --- in
the latter case it would be square]. The initial surface is at
$\eta$ = $\pi/2$; it is located at a finite proper time prior to
any other event in the spacetime
--- it is not a part of conformal infinity\footnote{This is
indicated by the dashed line in the diagram.}. The latter is at
$\eta$ = $\pi$, and is of course spacelike; the spacetime has a
regular future spacelike conformal completion. Notice that only
one quarter of the events in this spacetime are ever visible to
the observer at $\chi$ = 0, so this spacetime is indeed physically
distinct from de Sitter spacetime. Notice too that conformal
infinity is \emph{connected} here, so the difficulties arising in
connection with the disconnected conformal boundary of de Sitter
spacetime are eliminated. [See \cite{kn:orbifold} for a discussion
of these issues.]

We use this example as the model for the formulation of the global
structure of a true string gas cosmology. Assuming eternal
acceleration, the spacetime should have a regular future spacelike
conformal completion; to the past [at a finite proper time] there
should be a boundary, so that the spacetime itself is a
manifold-with-boundary having a connected boundary, and its
conformal completion is a compact manifold-with-boundary having
two boundary components. One should not think of either boundary
as a physical object: the past boundary just represents the state
of the Universe at the self-dual radius.

In the string gas case, there will be one crucial difference from
the de Sitter case: the spatial sections will not be spherical,
but flat. For simplicity we can take them to be modelled on a
cubic torus. Such spacetimes are locally but not globally
isotropic, so strictly speaking they cannot be represented on a
Penrose diagram; however, we can inscribe any given cube in a
sphere, and use these spheres to represent the corresponding cube.
With this understanding, we can draw the Penrose diagram. Since
the extent of conformal time is finite for an asymptotically
locally de Sitter spacetime, the diagram is rectangular in shape.
[If one embeds it in the Penrose diagram for Minkowski space,
which can be done since all of the spacetimes here are locally
conformally flat, the ``rectangle" will bend down slightly at the
top and bend inward on the right hand side, but of course these
details do not affect anything we shall say.] The precise aspect
ratio of the diagram, however, is not fixed classically. In the
case of de Sitter spacetime, the shape is fixed by the fact that
the size of the minimal three-sphere is determined by the value of
the cosmological constant. Here, because the size of a torus is
not related to its curvature, there is no such relation.

What \emph{does} fix the aspect ratio? This can only be a matter
of speculation at the moment, but some hints may be gleaned from
the recent work of Bousso \cite{kn:bousso3} on the difficulties
which arise when one attempts to identify asymptotic observables
in cosmology. As is well known, the existence of cosmological
horizons in the asymptotically locally de Sitter case means that a
certain proportion of the events in spacetime are forever
invisible to an inertial observer, and this argues against the
existence of asymptotic observables such as an S-matrix --- though
Bousso argues that other observables may be definable in certain
cases. In the de Sitter case, shown in Figure 2, the proportion of
invisible events is 3/4. Notice however that the proportion would
be reduced to 1/2 if we had adopted the $\bbr$P$^3$ interpretation
of de Sitter spacetime, since the diagram is square in that case.
In the string gas case, we can do even better, in fact arbitrarily
better: by assuming that the rectangle is ``tall and thin"
\cite{kn:smash}\cite{kn:leblond}, we can reduce the fraction of
permanently invisible events as much as we wish, since the top
right-hand corner of the diagram will have an arbitrarily small
relative area in this case. Again, Bousso points out that even in
a \emph{decelerating} FRW model with \emph{non-compact} flat
spatial sections, there is at all times an infinitely large
spatial region outside an observer's ``causal diamond", so that
the observer is permanently ignorant of an infinite amount of
information stored there. This problem is solved very simply in
the string gas case if the Penrose diagram is tall and thin: at
all times the fraction of a spatial section contained in a causal
diamond is non-zero, and, if the observer waits for a sufficiently
long [but always finite] time, he will see an entire spatial
section. He can then predict the fate of all objects in the
Universe, including those ultimately destined to disappear beyond
the horizon. In short, Bousso's arguments are in fact strong a
priori arguments against infinite spatial sections, and in favour
of the toral spatial structure assumed in string gas cosmology.
They also suggest strongly that the Penrose diagram should be tall
and thin.

An interesting possible shape for the Penrose diagram is given in
Figure 3. The upper dot represents the present time, under the
generally [but not universally \cite{kn:luminet}] accepted
assumption that the compactness of the spatial sections is not
currently observable \cite{kn:cornish}; the horizontal line below
it represents decoupling. The region below this horizontal line
represents the combined eras of NEC violation, inflation,
radiation dominance, and so on. Note that, in the case
illustrated, an observer in the early Universe [represented by the
lower dot] could detect the spatial compactness, in the sense that
the same object is ``visible" in two opposite directions. This
feat will \emph{eventually} be possible again, if we wait long
enough.
\begin{figure}[!h]
\centering
\includegraphics[width=0.2\textwidth]{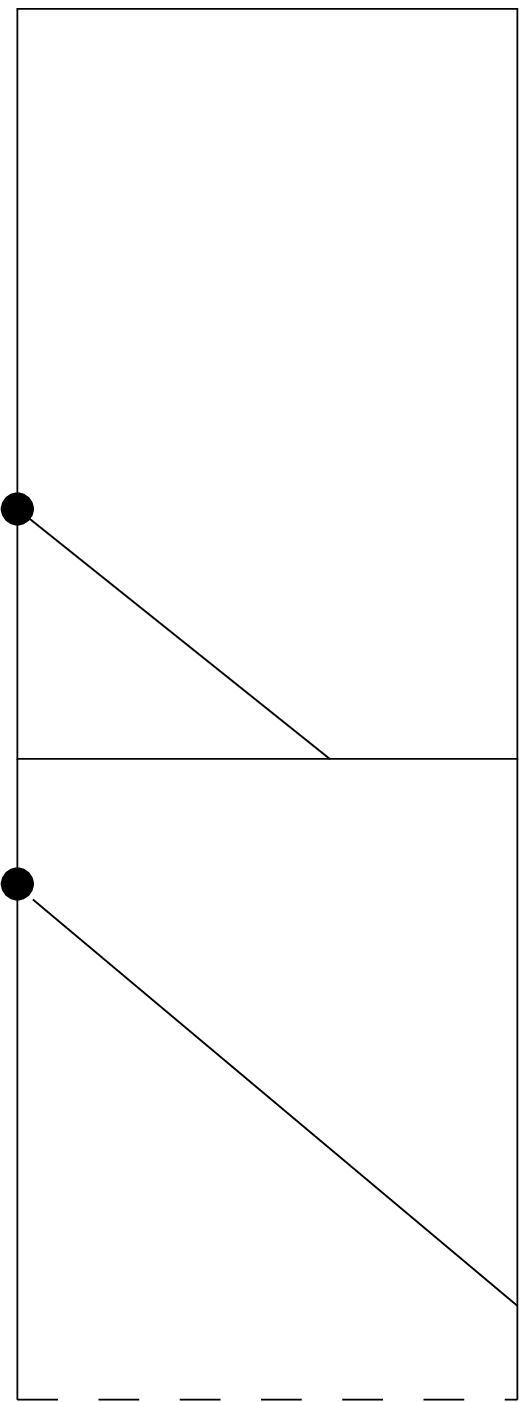}
\caption{Penrose diagram of a string gas cosmology}
\end{figure}

We have stressed that the past boundary should not be regarded as
a physical object: nothing ``special" should happen there. By this
we mean that T-duality should not obstruct attempts to probe
distances below the self-dual radius, but rather should reveal
them to be probes of larger distance scales --- this should be the
manner in which string gas cosmology \emph{smoothly} resolves
cosmological singularities. We can represent this in a formal way
as follows: the embedding of the initial surface should be such
that it is possible to glue an exact but ``upside-down" copy of
the Penrose diagram underneath the initial spacelike surface, in
such a way that the transition is smooth.

The resulting Penrose diagram can be obtained simply by taking one
like Figure 3 and ``doubling" it by flipping it around the dashed
line. It will \emph{formally resemble} that of a ``bounce"
cosmology, in the sense that there are components of conformal
infinity both at the bottom and at the top. However, the cosmology
in fact \emph{does not bounce}: time increases away from the
dashed line in \emph{both} directions\footnote{This formal
resemblance to a bounce cosmology is what allows us to escape the
conclusions of \cite{kn:guth}, where contraction, even of this
formal kind, is disallowed by assumption.}.

We regard this construction as a mere formal implementation of the
idea that T-duality smoothly resolves cosmological singularities.
The lower half of the ``doubled" spacetime is really identical to
the upper half; we have simply ``unwrapped" an orbifold. However,
there is an interesting analogy between this formal procedure and
the global spacetime structure proposed by Carroll and Chen, in
\cite{kn:chen}, as part of their theory of the arrow of time.
There, too, one has a distinguished surface such that time
increases away from it in both directions; in that case, however,
it is not essential that the lower region be identical to the
upper. It is nevertheless possible that true or effective NEC
violations are generic also in the Carroll-Chen
spacetime\footnote{The Carroll-Chen spacetime does \emph{not}
however have a regular future spacelike conformal completion,
since the structure never completely settles down to a de
Sitter-like state.}.

Although the resemblance of this model to bouncing cosmologies is
merely formal [so that, for example, we do not need to be
concerned about the propagation of perturbations through the
bounce], it nevertheless raises the same question that arises in
connection with bounces: if the spatial sections are flat, the
result could be an effective violation of all of the classical
energy conditions. However, as this argument is based on assuming
the precise form of the FRW metric [in particular, the exact
flatness of the spatial sections at all times] and on the exact
validity of the standard Friedmann equation, it is far from clear
that it can be regarded as conclusive. We shall now show that the
same conclusion can, however, be reached \emph{without} making
these assumptions.

First we remind the reader of the statement of the \emph{Null
Energy Condition:} this is simply the statement that for every
null tangent vector k$^\mu$ at every point of spacetime, the
[total] stress-energy-momentum tensor T$_{\mu\nu}$ should satisfy
\begin{equation}\label{eq:A}
\mathrm{T}_{\mu\nu}\,\mathrm{k}^\mu\,\mathrm{k}^\nu\;\geq\;0.
\end{equation}
In the cosmological context, this just means that any observer of
a cosmological spacetime with total energy density $\rho$ and
pressure p should find that
\begin{equation}\label{eq:B}
\rho\;\;+\;\;\mathrm{p}\;\;\geq\;\;0
\end{equation}
is satisfied everywhere.

This is a very mild energy condition: it is implied by both the
Weak and the Strong Energy Conditions, and it is satisfied both by
de Sitter and by anti-de Sitter spacetimes. There is evidence
\cite{kn:NEC} that it is essential for the functioning of the
Covariant Entropy Bound \cite{kn:bousso2}, and severe instability
problems can result if it is violated for ordinary matter fields.
However, recall from the Introduction the distinction between
``true" NEC violation [of the kind discussed in \cite{kn:carroll}]
and ``effective" NEC violation, which arises from including all
corrections to the Einstein field equations in the
stress-energy-momentum tensor. In this case, we will find that the
Universe evolves in a way which suggests that the NEC has been
violated, even if this is not actually the case. To see how to
detect this situation, we introduce the \emph{Null Ricci
Condition} [NRC], which requires that the Ricci tensor should
satisfy
\begin{equation}\label{eq:AA}
\mathrm{R}_{\mu\nu}\,\mathrm{k}^\mu\,\mathrm{k}^\nu\;\geq\;0.
\end{equation}
for every \emph{null} vector k$^\mu$. Clearly the NRC is exactly
equivalent to the NEC if the Einstein equations hold exactly. In
other cases, however, a violation of the NRC means that the
spacetime geometry is of the kind that would cause us to deduce
NEC violation if we used the Einstein equations to try to compute
the matter content from the geometry --- which is precisely what
is done, for example, when observers of supernovae use the data to
deduce apparent values of the dark energy equation-of-state
parameter below $-$1. That is, NRC violation is the formal
interpretation of ``effective" NEC violation.

We can now state the Andersson-Galloway singularity theorem
\cite{kn:andergall}:

\bigskip
\noindent THEOREM [Andersson-Galloway]: \emph{Let M$_4$ be a
globally hyperbolic spacetime with a regular future and past
spacelike conformal completion. If the Null Ricci Condition is
satisfied, and if the fundamental group of the Cauchy surfaces is
infinite, then M$_4$ can be neither future nor past asymptotically
simple.}
\bigskip

In physical language: if a [past and future] asymptotically
locally de Sitter spacetime is globally hyperbolic, has Cauchy
surfaces with an infinite fundamental group, and satisfies the NRC
[that is, it effectively satisfies the NEC], then it \emph{must}
be singular to the past and the future. In the spacetime pictured
in Figure 1, the manifold is Einstein and so the NRC is satisfied.
The spatial sections have the topology of the connected sum
$\bbr$P$^3$ $\#$ $\bbr$P$^3$, which has an infinite fundamental
group [it has the structure of a non-trivial U(1) bundle over
$\bbr$P$^2$]; thus the spacetime is forced to be singular by the
topology of its spatial sections. Conversely, if we know that a
spacetime with spatial sections of this kind is completely
non-singular, then we can say that \emph{the topology of the
spatial sections forces the spacetime to violate the NRC}, that
is, it enforces an effective violation of the NEC. Let us now see
in detail how the Andersson-Galloway theorem applies to string gas
cosmology.

Take a spacetime of the kind pictured in Figure 3, and ``double"
it, in the way we have described. This ``doubled" spacetime will
have a regular future and past spacelike conformal completion if
the original spacetime had a regular future spacelike conformal
completion. Similarly, the doubled space will be future and past
asymptotically simple if the original spacetime was free of
singularities. If the original spacetime is globally hyperbolic,
so will be the doubled one. Similarly, a string gas cosmology has
Cauchy surfaces which are either tori or finite quotients of tori,
and so the same is true of the doubled spacetime. \emph{All such
spaces have infinite fundamental groups}. Combining all of these
observations and applying the Andersson-Galloway theorem, we see
that, as claimed, the NEC must effectively fail in the doubled
spacetime, \emph{and so it must fail in the original spacetime}.

It is noteworthy that the only part of this discussion which does
not apply to de Sitter spacetime itself is the condition on the
fundamental group of the spatial sections [which is finite for de
Sitter spacetime and its relatives obtained by spatial topological
identifications \cite{kn:rp3}]. Thus de Sitter spacetime, which
satisfies the NEC at all times, escapes our conclusion \emph{only}
because it has the wrong \emph{topology} to be a string gas
cosmology. We see that the distinctive spatial topology of string
gas cosmology is at the heart of NEC violation, just as it is at
the heart of the main advantages of string gas cosmology.

As with the classical Hawking-Penrose singularity theorems, the
remarkable feature of the Andersson-Galloway theorem is that it
allows us to avoid any symmetry assumptions. It also does not
depend on the details of the Einstein equation. It is easy to
show, using the Friedmann equations, that FRW cosmologies with
\emph{exactly flat} spatial sections can only ``bounce" if the NEC
is [at least effectively] violated, just as it is easy to see that
all \emph{exact} FRW cosmologies satisfying the SEC have to be
singular. The point of the singularity theorems in both cases is
that we can arrive at the same conclusion no matter what happens
to the geometry of the spatial sections when the scale factor is
small, and without using the Friedmann equations. The relevant
behaviour in each case therefore cannot be avoided.

We conclude, then, that the avoidance of singularities in string
gas cosmologies entails certain costs: the NEC is [at least
effectively] violated in the earliest stages.

Just as a normal SEC-violating inflationary phase must come to an
end and be supplanted by radiation and matter-dominated phases, so
also the NEC-violating phase here must come to an end, making the
transition to the inflationary phase. [An analogous statement
holds true in the ``phantom inflation" model
\cite{kn:piaozhou}\cite{kn:piaozhang}.] That is, expressed in
terms of the equation-of-state parameter w, the w $<$ $-$1 phase
must be succeeded by a w $>$ $-$1 era, corresponding to the
inflaton. But this crossing of the ``phantom divide" is highly
non-trivial, since it can lead to various unphysical effects
\cite{kn:vikman}. Indeed, it seems that no model based on a single
scalar field can perform this feat. However, simple models with
\emph{two} scalar fields \cite{kn:hu}\cite{kn:doran} are capable
of crossing the divide\footnote{We stress however that these works
are concerned with crossing the divide in the opposite direction
to the one we are concerned with here.}. As Hu \cite{kn:hu}
stresses, such models can be regarded as a way of taking into
account any hidden degrees of freedom in the dark energy [see also
\cite{kn:cai}]. Since string gas cosmologies \emph{must} be able
to cross the divide, we conclude that they must be of this general
kind. It is therefore highly interesting that the recent proposal
of Biswas et al \cite{kn:mazumdar2} for obtaining inflation in
string gas cosmology does in fact involve two scalar fields, which
however are not of the same kind as those considered by Hu. It
would be of great interest to combine these sets of ideas, to
account for the pre-inflationary NEC-violating phase and its
transition, crossing the phantom divide to the inflationary era,
in the context of string gas cosmology.

Given that string gas cosmology must be able to cross the phantom
divide in one direction, it is easy to imagine that it may cross
it in the other --- that is, in relatively recent times, w may
have crossed from w $\geq$ $-$1 to w $<$ $-$1. This is in fact the
kind of crossing discussed specifically in \cite{kn:hu} and
\cite{kn:doran}. The question, motivated by persistent
observational suggestions that w may indeed currently be less than
$-$1 [and also by analyses showing that better fits to the data
\cite{kn:peri} are obtained by models which cross the phantom
divide in recent times], is whether string gas cosmologies can
sustain NEC violation not just briefly, in the extremely early
Universe, but also over cosmologically significant periods. We
shall now argue that the answer is ``no": there can be no second
crossing of the divide according to string theory.

\addtocounter{section}{1}
\section*{\large{3. NEC-Violating Cosmologies Unstable to Brane Pair Creation}}
In this section, we shall attempt to understand the constraints
imposed on an NEC-violating cosmology by embedding it in string
theory; this is reasonable, since string gas cosmologies are
intrinsically stringy. Of course, the whole problem of deriving de
Sitter spacetime from string theory is a formidable one. We have
nothing to contribute to this question here; we shall simply
assume that it can be done. Now de Sitter spacetime itself is not,
of course, a realistic cosmology. The task of understanding the
cosmic acceleration in string theory will only be complete after
we have studied the back-reaction of various kinds of matter,
\emph{including specifically stringy matter}, on the spacetime
geometry. In particular, the whole structure will be internally
consistent only if we can demonstrate that a de Sitter background
is not destabilized by stringy objects such as branes. As we shall
see, this is no trivial requirement.

When the AdS/CFT correspondence is generalized away from Anti-de
Sitter spacetime itself, there is a danger from various kinds of
instability, perturbative and non-perturbative \cite{kn:porrati}.
The particular effect with which we are concerned here involves
the production, by means of a ``Schwinger-like" pair-creation
process, of BPS brane-antibrane pairs. This intrinsically
non-perturbative process has been studied in depth in
\cite{kn:mich}\cite{kn:seiberg}\cite{kn:yau}\cite{kn:galloway}\cite{kn:maoz}
\cite{kn:buchel}\cite{kn:mcinnes}\cite{kn:answering}\cite{kn:maoz3}\cite{kn:porrati}\cite{kn:reallyflat},
and is well-understood in many cases, such as Einstein bulks with
positively or negatively curved conformal boundaries. In the
latter case, it is known that the pair-creation of branes is an
unstable process.

We shall consider (D $-$ 1)-branes in a space which generalizes
AdS$_{\mathrm{D} + 1}$ in some way. For the analysis of such
generalized geometries, see \cite{kn:wittenads}; note that, as in
this reference and in all of those given above, all of our
discussions of brane physics refer to the Euclidean domain, so
``AdS" means hyperbolic space with a specific kind of foliation.
The nucleation of brane-antibrane pairs in the Euclidean version
will of course translate back to the Lorentzian domain in various
ways, depending on the way in which the analytic continuation is
done.

Although this is not always emphasised, asymptotically hyperbolic
spaces can in fact be continued back to asymptotically locally de
Sitter spacetimes [in the signature ($+\;-\;-\;-$), which is why
we have used that signature]. For example, if the hyperbolic space
H$^4$ is globally foliated by copies of H$^3$, so that its metric
is
\begin{eqnarray}\label{eq:AARDVARK}
g(\mathrm{H}^4)\; =\; \m{dt^2}\;+\;\m{cosh^2(t/L)}[\mathrm{dr^2}\;
+\; \mathrm{L^2\,sinh^2(r/L)}\{\mathrm{d}\theta^2 \;+\;
\mathrm{sin}^2(\theta)\,\mathrm{d}\phi^2\}],
\end{eqnarray}
[see for example \cite{kn:randall}] then the continuation r
$\rightarrow$ iL$\chi$ yields
\begin{eqnarray}\label{eq:ANACONDA}
g(\mathrm{dS_4})\; =\;
\m{dt^2}\;-\;\m{cosh^2(t/L)\,L^2}\,[\mathrm{d\chi^2\; +\;
sin^2(\chi)}\{\mathrm{d}\theta^2 \;+\;
\mathrm{sin}^2(\theta)\,\mathrm{d}\phi^2\}],
\end{eqnarray}
and this is precisely the global de Sitter metric [equation
(\ref{eq:AAA}), in different coordinates]. It has the same
curvature [ = $-$1/L$^2$] as hyperbolic space, and also exactly
the same isometry group [the orthogonal group O(1, 4)]. It is
reasonable to assume that an asymptotically locally de Sitter
spacetime with a radically unstable Euclidean version must itself
be unstable. We shall not need to discuss the precise form this
instability might take\footnote{The obvious guess is that it
involves ``S-branes" \cite{kn:strominger}\cite{kn:jones}.}.

Returning to the Euclidean case, the least-understood situation
involves bulk manifolds which are not Einstein, and conformal
boundaries which are scalar flat; even in the case of an Einstein
bulk, the leading term in the action expansion given in
\cite{kn:seiberg} vanishes, and the outcome is determined by very
complex higher-order terms. [See \cite{kn:porrati} for a
discussion of these terms in this case.] \emph{But this
least-understood case is, of course, precisely the case of most
interest in cosmology}, since cosmological spacetimes are almost
never Einstein, and since the flat spatial sections of a physical
FRW cosmology induce a flat structure at infinity. Thus it is not
clear whether the kind of stringy instability we have been
discussing can be avoided in the cosmological context; this is a
fundamental question for any string cosmology.

It will be shown elsewhere \cite{kn:stable} that if a cosmological
model is asymptotically locally de Sitter towards the future, has
flat spatial sections, and contains matter \emph{satisfying} the
NEC, then this non-perturbative instability cannot
arise\footnote{These spacetimes are singular, by the
Andersson-Galloway theorem.}. The question now is whether this
satisfactory result is valid also in the case of string gas
cosmology.

The brane action is, for a BPS brane,
\begin{equation}\label{eq:C}
\mathrm{S} \;=\;
\mathrm{T}(\mathrm{A}\;-\;{{\mathrm{D}}\over{\mathrm{L}}}\,\mathrm{V}),
\end{equation}
where the bulk is (D+1)-dimensional, T is the tension, A is the
area, V the volume enclosed, and L is a constant length related to
the asymptotic cosmological constant $\Lambda_{\infty}$ by
\begin{equation}\label{eq:CC}
\m{\Lambda_{\infty}\;=\;-\,D/L^2}.
\end{equation}
One sees at once that there is a danger that the volume term might
dominate in some region of the bulk, in which case the nucleation
of a brane-antibrane pair [a non-perturbative effect, since a
barrier has to be overcome] would be an unstable process, since
one could always move one member of the pair to the region where S
is negative. [See \cite{kn:maoz} for explicit examples where this
happens, not necessarily near infinity.]

The essential point here is that the question as to which term
dominates is a purely \emph{geometric} one. The process does not
``care" how we obtain a specific geometry --- whether, for
example, a specific shape is the result of ``true" or ``effective"
violations of some energy condition. Once the metric is fixed, the
question of the positivity of the brane action is determined.
Henceforth, therefore, everything we say will hold for
\emph{either} kind of NEC violation.

In the specific case of the string gas geometry, as pictured in
Figure 3, it is clear that the area term will dominate near to the
initial surface; this is true whether or not the NEC is violated
in that region. Thus the function S will be positive initially.
The question is whether it \emph{remains} positive.

Unfortunately there is no general theory regarding this question
in the case of interest to us. Even in the case where the bulk is
Einstein, the question of non-perturbative stability when the
conformal boundary is \emph{flat} is a subtle one, depending on
certain constants [the ``regularized area and volume"
\cite{kn:porrati}] which one does not know how to evaluate in
general. We can however investigate concrete examples, provided
that we know the metric exactly, and we shall do this below. We
shall first show very simply that there does exist a large family
of NEC-violating cosmologies which \emph{are} unstable against
brane pair creation. The issue is thus a very real one.

We shall be interested in cosmological models with metrics of the
general form
\begin{equation}\label{eq:D}
g^- \;=\; \m{+\,\m{dt}^2\; -\; \m{A^2\;a(t)}^2[d\theta_1^2 \;+\;
d\theta_2^2 \;+\; d\theta_3^2]},
\end{equation}
where the minus sign reminds us that we are in Lorentzian
signature, where a(t) is the scale factor, and where the spatial
coordinates are angular coordinates on a cubic torus with all
circumferences equal to 2$\pi$A on the initial surface [so that
a(t) = 1 there]. This metric will resemble the de Sitter metric
[in the flat slicing] at large t provided that the function a(t)
has the appropriate [exponential] form there. That is, at large t
the spacetime will closely resemble a spacetime of \emph{constant
negative curvature}. Following our discussion of de Sitter
spacetime above, this means that the appropriate Euclidean version
of this metric is simply
\begin{equation}\label{eq:DONUT}
g^+ \;=\; \m{+\,\m{dt}^2\; +\; \m{A^2\;a(t)}^2[d\theta_1^2 \;+\;
d\theta_2^2 \;+\; d\theta_3^2]},
\end{equation}
since this metric approaches a metric of constant \emph{negative}
curvature for large t, that is, it is an asymptotically hyperbolic
space. If one likes to think in terms of analytic continuation,
one can move between the two versions by means of A $\rightarrow$
iA.

It will be useful to describe the matter content of our
NEC-violating models in the following way. We shall certainly want
the local spacetime geometry to become indistinguishable from that
of de Sitter space after the passage of a certain amount of time.
To keep this to the fore, we shall write the total density and
pressure as sums of the form
\begin{equation}\label{eq:DRAGON}
\m{\rho\;=\;{{3}\over{8\pi L^2}}\;+\;\rho_{\psi}},
\end{equation}
\begin{equation}\label{eq:DUNCE}
\m{p\;=\; {{-\,3}\over{8\pi L^2}}\;+\;p_{\psi}}.
\end{equation}
Obviously, it is always formally possible to write the energy
density and the pressure in this way; the reader is free to regard
this procedure as a mere formal device, or he may prefer to think
in terms of some kind of more-or-less exotic ``matter field"
[labelled $\psi$, with density $\rho_{\psi}$ and pressure
p$_{\psi}$] propagating on a de Sitter background with
cosmological constant $-$3/L$^2$, energy density 3/8$\pi$L$^2$,
and pressure $-\,3/8\pi$L$^2$. We do not necessarily regard this
field as ``matter" in the ordinary sense
---  it may simply parametrize the effects of higher-order
curvature terms, of quantum effects
\cite{kn:onemli1}\cite{kn:onemli2}, of brane-world physics, or
other such contributions arising from the embedding of the system
in higher-dimensional physics
\cite{kn:varun}\cite{kn:varun2}\cite{kn:nojiri}\cite{kn:tsujikawa}\cite{kn:kachru}.

We are interested in determining whether an NEC-violating
cosmology can be stable against brane pair-production. For clarity
let us assume that the NEC is always violated. This means that
$\rho\;+\;\m{p}$ [equivalently, $\rho_{\psi}\;+\;\m{p}_{\psi}$] is
negative. Since a suitable combination of components of Einstein's
equation for a spatially flat FRW cosmology yields
\begin{equation}\label{eq:DRUG}
\m{\dot{H}(t)\;=\;-\,4\pi\,(\rho_{\psi}\;+\;p_{\psi})},
\end{equation}
where H is the Hubble parameter and the dot denotes a cosmic time
derivative, we see that H will always increase here.

Now take equation (\ref{eq:C}) and evaluate it in the case of the
Euclidean metric (\ref{eq:DONUT}), obtaining
\begin{equation}\label{eq:DONKEY}
\m{S(t)\;=\;T[8\pi^3A^3a(t)^3\;-\;{{24\pi^3A^3}\over{L}}\int_0^t
a(\tau)^3\,d\tau ]}.
\end{equation}
This expression is hard to handle, but the derivative is more
informative: one has
\begin{equation}\label{eq:DRAKE}
\m{\dot{S}(t)\;=\;24\pi^3\,T\,A^3\,a(t)^3\,[H\;-\;{{1}\over{L}}]}.
\end{equation}
In our case H(0) = 0, and its asymptotic value in an
asymptotically de Sitter spacetime is 1/L; on the other hand, we
saw above that H always increases in a spacetime which violates
the NEC. It follows that H is always smaller than 1/L, and hence
that S(t) is a function which steadily decreases from its initial
positive value of $8\pi^3\,\m{T\,A^3}$. It could still be
asymptotic to a positive value, however.

The right side of equation (\ref{eq:DRAKE}) contains a product of
a(t)$^3$, which increases without limit in an asymptotically de
Sitter spacetime, with H $-$ 1/L, which tends to zero. Simple
calculus now implies that
\begin{equation}\label{eq:DOG}
\m{\lim_{t \rightarrow
\infty}\dot{S}(t)\;=\;-\,8\pi^3\,T\,A^3\,\lim_{t \rightarrow
\infty}\,[a(t)^3\,\dot{H}/H]},
\end{equation}
and now equation (\ref{eq:DRUG}) gives
\begin{equation}\label{eq:DENMARK}
\m{\lim_{t \rightarrow
\infty}\dot{S}(t)\;=\;32\pi^4\,T\,A^3\,L\,\lim_{t \rightarrow
\infty}\,[a(t)^3\,(\rho\;+\;p)]},
\end{equation}
which can also be usefully written as
\begin{equation}\label{eq:DARK}
\m{\lim_{t \rightarrow
\infty}\dot{S}(t)\;=\;32\pi^4\,T\,A^3\,L\,\lim_{t \rightarrow
\infty}\,[a(t)^3\,(\rho_{\psi}\;+\;p_{\psi})]}.
\end{equation}
Since $\rho\;+\;\m{p}$ and $\rho_{\psi}\;+\;\m{p}_{\psi}$ are
negative, it is clear that if the limit on the right is a non-zero
constant or diverges, then the derivative $\m{\dot{S}(t)}$ either
approaches a constant negative value or diverges to
\emph{negative} infinity; in either case S(t) will not only be
eventually negative but also unbounded below. \emph{We conclude
that these cosmologies will definitely be unstable to brane pair
production if $\rho_{\psi}\;+\;\m{p}_{\psi}$ tends to zero more
slowly than a(t)$^{-\,3}$}.

We have derived this conclusion without making any assumptions
other than that the NEC is violated [in either sense]. We shall
now see that far stronger conclusions can be reached in the
context of specific spacetime geometries.

\addtocounter{section}{1}
\section*{\large{4. Brane Pair Production: Constant EOS Parameter}}
In the previous section we saw that the NEC-violating matter has
to dilute rapidly in order to avoid instability due to brane pair
production. The condition we derived was, however, merely a
\emph{necessary} one for stability: it is not clear that it is
sufficient. With the help of specific examples, we shall now argue
that it is \emph{not} sufficient. Ultimately the real question is
whether it is at all possible to avoid this problem, and, if so,
at what cost.

In order to proceed, we need to make some assumptions about
$\rho_{\psi}$ and p$_{\psi}$; these will be as general and as
conservative as possible.

In early discussions of phantom cosmology, concern was often
expressed that the phantom energy might propagate outside the
light-cone. This is understandable: in phantom cosmologies one can
always find an observer who observes a spacelike energy-momentum
vector for the phantom matter. However, the situation is not as
simple as it seems, because there is more than one way to build a
spacelike energy-momentum vector.

We shall distinguish between two cases, which we call
``semi-exotic" and ``fully exotic". In the first case, the
energy-momentum vector \emph{of the $\psi$ field alone} is
timelike but past-pointing. This kind of matter is ``semi-exotic"
in the sense that while a past-pointing energy-momentum vector is
of course unorthodox, at least there can be no question of the
energy of the $\psi$ matter propagating outside the light cone.
[Of course we do not need to think of energy ``propagating
backwards in time" here, any more than one speaks of ordinary
energy ``propagating" forwards in time.] In terms of $\rho_{\psi}$
and p$_{\psi}$, we are requiring that p$_{\psi}$ should be related
to $\rho_{\psi}$ [which is negative] by
\begin{equation}\label{eq:G}
\m{p_{\psi}\;=\;w_{\psi}\,\rho_{\psi}},
\end{equation}
where w$_{\psi}$, the equation-of-state parameter for $\psi$
alone, satisfies $\m{-1\;<\;w_{\psi}\;<\;1}$ so that the
energy-momentum vector for $\psi$ alone remains timelike. The
point now is that the energy-momentum four-vector of the entire
system will be spacelike if we insert a small amount of this
``semi-exotic" field into de Sitter spacetime, simply because the
energy-momentum four-vector of the constant dark energy is null
and future-pointing, and the sum of a ``small" past-pointing
timelike four-vector with a future-pointing null vector will be
spacelike. This achieves a violation of the NEC [corresponding to
a total energy-momentum four-vector which is spacelike]
\emph{without} any propagation of energy outside the light-cone,
since the kind of dark energy represented by a cosmological
constant cannot ``propagate" --- its energy density is rigidly
constant in space and time. By contrast, if the $\psi$ field
itself [assumed not to exhibit this ``rigidity"] already has a
spacelike energy-momentum vector, then we are in the ``fully
exotic" case, and causality may be a real issue.

We assume for simplicity that w$_{\psi}$ does not vary with time.
With this assumption one can solve the Einstein equation for a
metric of the form (\ref{eq:D}), obtaining the exact solution
\cite{kn:smash}:
\begin{equation}\label{eq:E}
g^-(\gamma,\,\m{A,\,L) \;=\; +\,dt^2\; -\;
A^2\;cosh^{({{4}\over{\gamma}})}({{\gamma
t}\over{2L}})[d\theta_1^2 \;+\; d\theta_2^2 \;+\; d\theta_3^2]},
\end{equation}
where $\gamma$ is a constant given by
\begin{equation}\label{eq:EE}
\gamma\;=\;3\,[1\;+\;\m{w_{\psi}]}.
\end{equation}
Clearly this geometry can be interpreted in the way we have
suggested: that is, we can regard the t $<$ 0 part of the geometry
as a mere copy of the t $>$ 0 part; the surface t = 0 is the
orbifold surface at which the Universe begins. The conformal
boundary is connected, compact, spacelike, and \emph{flat}, that
is, the conformal structure can be represented by a flat metric.

The requirement that the energy-momentum vector of the $\psi$
field should be timelike is here expressed by
\begin{equation}\label{eq:EEE}
0\;<\;\gamma\;<\;6,
\end{equation}
so the value 6 marks the boundary between the semi-exotic and
fully exotic cases. This metric is asymptotically locally de
Sitter towards the future: the scalar curvature is given by
\begin{equation}\label{eq:F}
\m{R}(g^-(\gamma,\,\m{A,\,L)) \;=\;
-\,{{12}\over{L^2}}\;+\;{{3}\over{L^2}}\,(4\;-\;\gamma)\,sech^2({{\gamma
t}\over{2L}})},
\end{equation}
with an asymptotic value of $-$ 12/L$^2$, in agreement with the de
Sitter value from equation (\ref{eq:AAA}) [with signature as
given]. The NEC is violated at all times in this spacetime: one
has
\begin{equation}\label{eq:GALOSHES}
\m{\rho_{\psi}\;+\;p_{\psi}\;=\;{{-\,\gamma}\over{8\pi
L^2}}\,sech^2({{\gamma t}\over{2L}})}.
\end{equation}
Notice that the NEC-violating matter is diluted towards t =
$\infty$, and is concentrated near t = 0. In fact we have
\begin{equation}\label{eq:GOPHER}
|\m{\rho_{\psi}\;+\;p_{\psi}|_{Max}\;=\;{{\gamma}\over{8\pi
L^2}}},
\end{equation}
so $\gamma$ measures the degree to which the NEC is violated; as
expected, the violation is less severe in the preferred,
``semi-exotic" case. On the other hand, the value of t when this
quantity falls to half of its maximal value, t$_{1/2}$, is given
by
\begin{equation}\label{eq:GADZOOKS}
\m{t_{1/2}\;=\;{{2\,L}\over{\gamma}}\,cosh^{-\,1}(\sqrt{2})},
\end{equation}
so we see that the NEC-violating ``matter" dies off away from t =
0 \emph{more slowly} when $\gamma$ is small. Large values of
$\gamma$ correspond to an NEC-violating phase which is intense but
short-lived. For an application involving lengthy intervals of
time, this suggests that \emph{small} values of $\gamma$ are
preferred.

The Penrose diagram of this spacetime is rectangular, with width
determined by A and height determined by L and $\gamma$: the
height is in fact
\begin{equation}\label{eq:GODZILLA}
\m{\int_0^\infty{{dt}\over{cosh^{({{2}\over{\gamma}})}({{\gamma
t}\over{2L}})}}},
\end{equation}
which is always larger than L, and which becomes arbitrarily large
for sufficiently \emph{small} $\gamma$. In view of our previous
discussion of ``tall and thin" Penrose diagrams, this is further
evidence that it would be best to remain in the ``semi-exotic"
regime of small values for $\gamma$. Unfortunately we shall now
see that this is not possible.

We can write
\begin{equation}\label{eq:H}
\m{\rho_{\psi}\;+\;p_{\psi}\;=\;{{-\,\gamma}\over{8\pi
L^2}}\,a(t)^{-\,\gamma}}.
\end{equation}
Comparing this with equation (\ref{eq:DARK}), we see that this
model will be unstable against the production of brane pairs if
$\gamma$ takes any value less than or equal to 3. Thus for example
equation (\ref{eq:E}) with $\gamma$ = 2,
\begin{equation}\label{eq:I}
g^-(2,\,\m{A,\,L) \;=\; +\,dt^2\; -\;
A^2\;cosh^2({{t}\over{L}})[d\theta_1^2 \;+\; d\theta_2^2 \;+\;
d\theta_3^2]},
\end{equation}
represents a spacetime which is strongly unstable against brane
pair creation, the brane action being unbounded below. \emph{This
is the first known explicit example of a spacetime which is
unstable in the Seiberg-Witten \cite{kn:seiberg} sense despite
\emph{not} having a negatively curved conformal boundary}. Notice
that it differs from de Sitter spacetime only in that the spatial
sections are flat and compact instead of spherical. The
introduction of NEC-violating matter into de Sitter spacetime
allows us to flatten its spatial sections, but at the cost of
rendering it strongly unstable.

While the general argument of the preceding section does not apply
to these spacetimes when $\gamma$ exceeds 3, it is intuitively
clear that many of them will also have action functions
S$_{\gamma}$(t) which become [and therefore remain] negative. In
fact, in this case we can settle this by a direct calculation.
Substituting a(t) from equation (\ref{eq:E}) into equation
(\ref{eq:DONKEY}), we have
\begin{equation}\label{eq:J}
\m{S_{\gamma}(t)\;=\;T[8\pi^3A^3cosh^{(6/\gamma)}({{\gamma
t}\over{2L}})\;-\;{{24\pi^3A^3}\over{L}}\int_0^t
cosh^{(6/\gamma)}({{\gamma \tau}\over{2L}})\,d\tau ]}.
\end{equation}
This can be evaluated explicitly in many cases of interest: for
example, in the ``borderline" case $\gamma$ = 3 [which we know to
be unstable] we have
\begin{equation}\label{eq:K}
\m{S_{3}(t)\;=\;4\pi^3A^3T[1\;+\;e^{-\,3t/L}\;-\;{{3t}\over{2L}}]},
\end{equation}
which is indeed unbounded below.

Since we understand the case $\gamma\;\leq\;3$, let us assume that
$\gamma\;>\;3/2$. Then for large t we can approximate
$\m{cosh^{(6/\gamma)}({{\gamma t}\over{2L}})}$ by the first two
terms in its Taylor expansion. Doing this, we find after extensive
simplifications that, approximately,
\begin{equation}\label{eq:L}
\m{S_{\gamma}(t)\;\approx\;2^{(3\;-\;{{6}\over{\gamma}})}\,\pi^3\,A^3\,T\,[{{6}\over{\gamma\;-\;3}}\,
e^{(3\;-\;\gamma)t/L}\;-\;{{(6\;-\;\gamma)(3\;+\;\gamma)}\over{\gamma(\gamma\;-\;3)}}
]},
\end{equation}
where we can exclude $\gamma = 3$ since we have already treated
that case exactly. In agreement with our general result, we find
that if $\gamma$ is slightly less than 3, then the exponential
term dominates and is negative, so S$_{\gamma}$(t) is unbounded
below. The new result is that, if $\gamma$ is larger than 3 but
[in accordance with (\ref{eq:EEE}) above] less than 6, then the
constant term dominates and is \emph{negative}. In the limit the
approximation becomes exact, and we have, for any $\gamma\;>\;3$,
\begin{equation}\label{eq:M}
\m{\lim_{t \rightarrow
\infty}{S_{\gamma}(t)}\;=\;-\,2^{(3\;-\;{{6}\over{\gamma}})}\,\pi^3\,A^3\,T\,
{{(6\;-\;\gamma)(3\;+\;\gamma)}\over{\gamma(\gamma\;-\;3)}}}.
\end{equation}
Thus in this case S$_{\gamma}$(t) will decline to a negative
value. At any point it can always be made smaller, by moving to
larger values of t; nevertheless, in this case it is bounded below
[by its limit]. We see however that the pair-production
instability is present for all of our ``semi-exotic" models.

It is of course true that the right side of (\ref{eq:M}) can be
made arbitrarily close to zero in magnitude by taking $\gamma$
sufficiently close to 6. Then S$_{\gamma}$(t) will still become
negative, but only at a huge value of t. In this way, one can
postpone the onset of instability until the remote future, at the
cost of a severe fine-tuning.

The alternative is to ignore the inequalities (\ref{eq:EEE}) and
allow the $\psi$ field to have a spacelike energy-momentum vector;
in this ``fully exotic" case, S$_{\gamma}$(t) certainly remains
positive everywhere, but one has to be concerned about the
possible propagation of energy outside the light cone.
Nevertheless, such possibilities have been seriously considered
\cite{kn:turok}, and we do not rule them out completely, at least
not in the very early Universe. In short, it \emph{is} possible to
keep the brane action positive everywhere, but only by means of
choosing rather extreme parameter values.

Interesting as all these possibilities may be, however, the key
point is this: if $\gamma$ is substantially larger than 3, then
our theory \emph{cannot account for any observations corresponding
to a value of w $<$ $-$1 at the present time}. For if $\gamma$ is
so large, we see from (\ref{eq:H}) that the NEC-violating field
decays away extremely rapidly with the growth of the Universe, far
more rapidly than either relativistic or non-relativistic matter,
and of course far faster than cosmological constant-style dark
energy, quintessence, or an inflaton. This is exactly what we want
in the very early Universe, to ensure a crossing of the phantom
divide so that inflation can proceed; but it is not what we want
if we are trying to account for purported observations of phantom
behaviour at the \emph{present} time. In fact, to account for such
observations without interfering [for example] with the standard
nucleosynthesis computations, we would need $\gamma$ to be very
small, far below 3, and we know that this is out of the question.

To summarize, we have found that string theory only tolerates NEC
violation if the deviation from a pure positive cosmological
constant decays extremely rapidly: specifically, it must
[according to equation (\ref{eq:H})] decay according to at least
the inverse sixth power of the scale factor. This is highly
desirable if we are describing the NEC-violating phase which,
according to string gas cosmology, must precede inflation
--- the NEC violation decays very rapidly and allows inflation to
proceed in the way described in \cite{kn:mazumdar1} or
\cite{kn:mazumdar2}. However, it also means that string theory
predicts that there can be no appreciable deviation of the dark
energy equation-of-state parameter below $-$1 at the present time.
This conclusion holds even if the NEC ``violation" is of the
``effective" kind.

\addtocounter{section}{1}
\section*{5. Specific Potentials}
Thus far, we have not tried to be more precise about the nature of
the $\psi$ field; we saw that this is not really necessary in
order to draw our conclusions. Now, however, that we know that NEC
violation is relevant only in the extremely early Universe, for a
very short time, we can allow ourselves to consider more specific
models for the $\psi$ field. As is well known \cite{kn:caldwell},
the simplest possible parametrization of it is as a scalar field
with a kinetic term of the ``wrong" sign. It is also well known
\cite{kn:carroll} that this particular kind of phantom model is
subject to quantum instabilities, but perhaps this is of less
concern for a model which does not attempt to violate the NEC over
a cosmologically significant period. In any case, we simply note
here that if we introduce an explicit $\psi$ field of this kind,
then the precise spacetime \cite{kn:smash} we discussed above
[with metric
 $g^-(\gamma,\,\m{A,\,L})$] can be obtained from the surprisingly simple potential
\begin{equation}\label{eq:N}
\m{V_{\gamma}(\psi)\; = \; {{\gamma \;-\;6}\over{16\pi
L^2}}\,cos^2(\sqrt{2\pi \gamma}\;\psi)}.
\end{equation}
Bear in mind here that this potential is to be superimposed on a
de Sitter background, so the claim is that
$g^-(\gamma,\,\m{A,\,L})$ solves the Friedmann equation when the
total density is the sum of 3/8$\pi$L$^2$, the negative kinetic
term for $\psi$, and V$_{\gamma}(\psi$). One can also verify using
the field equation for $\psi$ that, with this potential,
w$_{\psi}$ is indeed constant, as we assumed. [Notice that
V$_{\gamma}(\psi$) is positive when $\gamma$ exceeds 6, and this
is indeed precisely the condition for the magnitude of the
pressure, ${{1}\over{2}}\dot{\psi}^2\;+\;$V$_{\gamma}(\psi)$ for a
scalar with a reversed kinetic term, to exceed the magnitude of
the density, ${{1}\over{2}}\dot{\psi}^2\;-\;$V$_{\gamma}(\psi)$.]
A two-scalar model, consisting of a $\psi$-field with a potential
similar to (\ref{eq:N}), together with a more conventional
inflaton field, could give a simple quantitative description of
the crossing of the phantom divide in the early Universe. [See
\cite{kn:veneziano} for a recent discussion of related models in
the context of bouncing cosmologies.]

The reader may object that our strongest conclusions depend on the
assumption that w$_{\psi}$ is independent of time --- this
corresponds to the specific potential (\ref{eq:N}) and the exact
metric (\ref{eq:E}). This is of course true, though it is hard to
believe that a relaxation of this assumption can materially change
our conclusions. Nevertheless it is desirable to check our results
with an exact phantom model having a non-constant w$_{\psi}$. In
fact, an exact phantom cosmological solution has recently become
available \cite{kn:arafeva} in which w$_{\psi}$ is not constant,
and it is very interesting to examine that solution from this
point of view.

Aref'eva, Koshelev, and Vernov consider a potential of the form
\begin{equation}\label{eq:O}
\m{V_{AKV}(\psi)\; = \;
{{1}\over{2\,L^2}}\,[1\;-\;\psi^2]^2\;+\;{{1}\over{12\,M_p^2\,L^2}}\,\psi^2[3\;-\;\psi^2]^2},
\end{equation}
where L is a constant length which sets the scale of the potential
at the initial time, and where M$_{\m{p}}$ is a gravitational
coupling constant. [Note that small M$_{\m{p}}$ corresponds to
strong coupling.] This theory is a sort of truncation of the one
we have been considering above; it has several remarkable
properties, including the existence of an exact solution for a
metric of the form (\ref{eq:D}), with the scale function given by
\begin{equation}\label{eq:P}
\m{a(t)_{AKV}\;=\;cosh^{1/(3M^2_p)}(t/L)\,exp[{{1}\over{12M^2_p}}\,tanh^2(t/L)]}.
\end{equation}
This spacetime can be interpreted as a string gas cosmology, with
the initial surface at t = 0. It is asymptotically locally de
Sitter with asymptotic cosmological constant
\begin{equation}\label{eq:Q}
\m{\Lambda_{AKV\infty}\;=\;-\,1/(3M^4_p\,L^2)}.
\end{equation}
The total energy density and pressure are given \cite{kn:arafeva}
by
\begin{equation}\label{eq:R}
\m{\rho_{AKV}\;=\;{{1}\over{12\,M^2_pL^2}}\,tanh^2(t/L)\,[3\;-\;tanh^2(t/L)]^2}
\end{equation}
\begin{equation}\label{eq:S}
\m{p_{AKV}\;=\;-\,{{1}\over{L^2}}\,sech^4(t/L)\;-\;{{1}\over{12\,M^2_pL^2}}\,tanh^2(t/L)\,[3\;-\;tanh^2(t/L)]^2}.
\end{equation}
These expressions split very naturally according to the pattern of
(\ref{eq:DRAGON}) and (\ref{eq:DUNCE}); we can regard the phantom
matter here as a combination of a de Sitter cosmological constant
with a negative-density field:
\begin{equation}\label{eq:T}
\m{\rho_{AKV}\;=\;{{1}\over{3M^2_pL^2}}\;-\;
        {{1}\over{12\,M^2_pL^2}}\,sech^4(t/L)\,[3\;+\;sech^2(t/L)]}
\end{equation}
\begin{equation}\label{eq:U}
\m{p_{AKV}\;=\;-\,{{1}\over{3M^2_pL^2}}\;+\;sech^4(t/L)[-\,{{1}\over{L^2}}\;+\;{{1}\over{12\,M^2_pL^2}}\,[3\;+\;sech^2(t/L)]]}.
\end{equation}
Hence, in our notation, the equation-of-state parameter of the
negative-density field is
\begin{equation}\label{eq:V}
\m{w_{\psi\,AKV}\;=\;-\,1\;+\;{{12M_p^2}\over{3\;+\;sech^2(t/L)}}},
\end{equation}
and we see at once that $\m{w_{\psi\,AKV}}$ does indeed vary with
time, increasing steadily from its minimum value at t = 0 to its
maximum at infinity,
\begin{equation}\label{eq:W}
\m{w_{\psi\,AKV}(max)\;=\;-\,1\;+\;4\,M_p^2}.
\end{equation}
We see from (\ref{eq:V}) that $\m{w_{\psi\,AKV}}$ never falls
below $-$1, approaching it only in the limit where M$_{\m{p}}$
tends to zero
--- note that this is the limit of \emph{strong} coupling of the
phantom to gravity, but it is also the limit of small asymptotic
cosmological constant. If the theory is to remain ``semi-exotic",
that is, for the energy-momentum four-vector of the $\psi$ field
is to remain timelike, then $\m{w_{\psi\,AKV}(max)}$ must not
exceed unity, and so we must have
\begin{equation}\label{eq:X}
\m{M^2_p\; < \;1/2};
\end{equation}
that is, the coupling has to be at least this strong in the
semi-exotic case.

Now let us investigate the stability of these cosmologies against
brane pair creation. Substituting (\ref{eq:P}), (\ref{eq:T}) and
(\ref{eq:U}) into (\ref{eq:DENMARK}) we find that the derivative
of the brane action in this case satisfies
\begin{equation}\label{eq:Y}
\m{\lim_{t \rightarrow \infty}\dot{S}_{AKV}(M_p^2\; ;
\;t)\;=\;-\,32\pi^4\,T\,A^3\,L^{-1}\,exp({{1}\over{4M_p^2}})\,\lim_{t
\rightarrow \infty}\,cosh^{[M_p^{-2}\;-\;4]}(t/L)},
\end{equation}
so clearly we must have
\begin{equation}\label{eq:Z}
\m{M^2_p\; > \;1/4}
\end{equation}
if the brane action itself, S$_{\m{AKV}}(\m{M_p^2\;;\;t}$), is to
have any hope of remaining positive.
\begin{figure}[!h]
\centering
\includegraphics[width=0.7\textwidth]{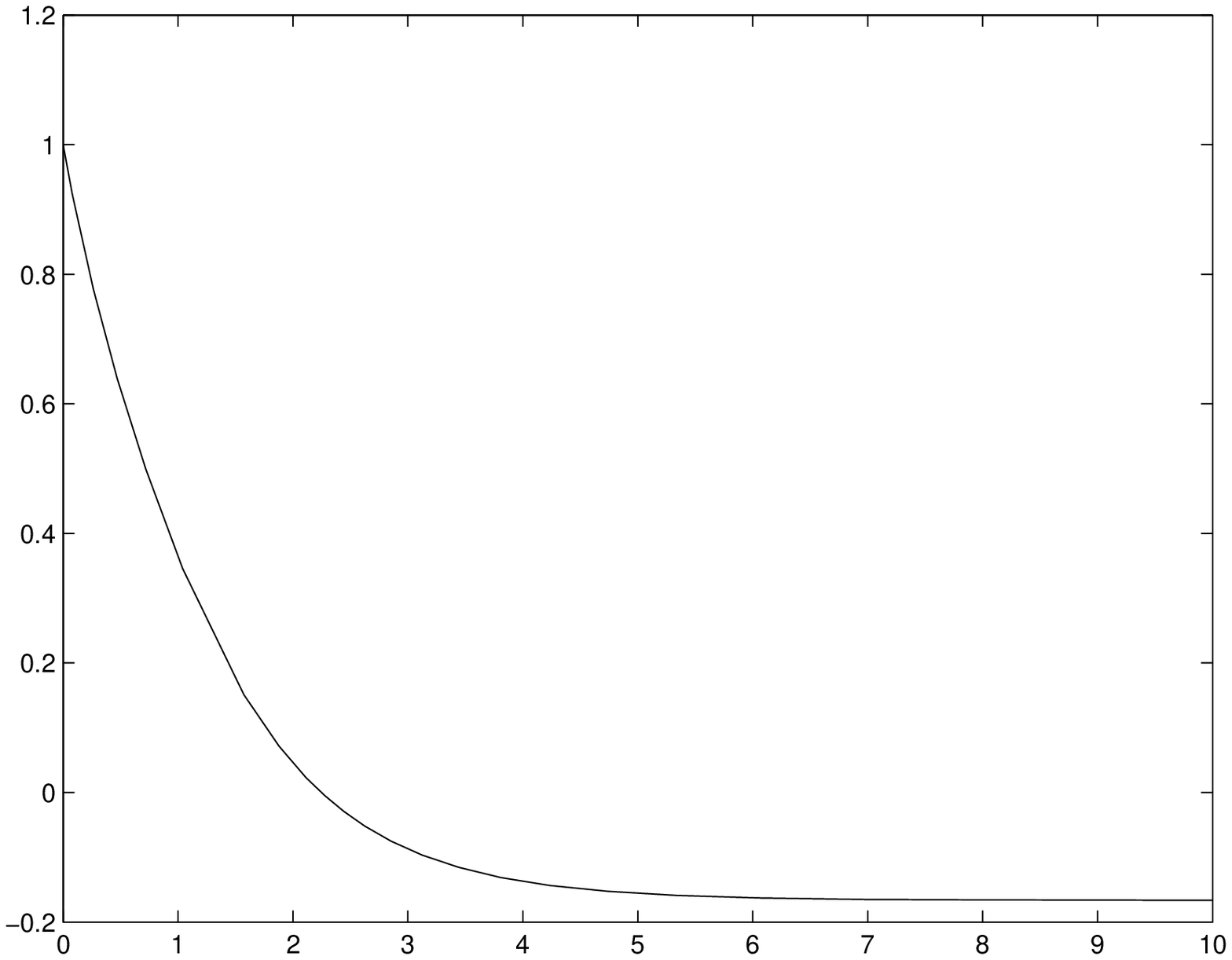}
\caption{The action S$_{\mathrm{AKV}}(1/3\;;\;\m{t})$.}
\end{figure}

In short, M$_{\m{p}}^2$ must in any case be at least 1/4; between
1/4 and 1/2 the non-de Sitter contribution to the energy-momentum
vector will be timelike; beyond 1/2, the contribution will be
spacelike. While we would therefore prefer values corresponding to
the ``semi-exotic" regime, our experience in the preceding section
leads us to expect that, while values between 1/4 and 1/2 will
lead to an action which is bounded below, the action will
nevertheless become negative in this case.

The brane action in this case is
\begin{eqnarray}\label{eq:DEMON}
\m{S_{AKV}(M_p^2\;;\;t)} & = & \m{8\pi^3A^3T \Big\{cosh^{1/(M^2_p)}(t/L)\,exp[{{1}\over{4M^2_p}}\,tanh^2(t/L)]} \nonumber\\
& & -\;\m{{{3}\over{L}}\int_0^t
cosh^{1/(M^2_p)}(\tau/L)\,exp[{{1}\over{4M^2_p}}\,tanh^2(\tau/L)]\,d\tau
\Big\}},
\end{eqnarray}
which of course has to be studied numerically. For simplicity we
choose the parameters so that L = 8$\pi^3$A$^3$T = 1, so that
S$_{\m{AKV}}(\m{M_p^2\;;\;0}$) = 1 for all choices of
M$_{\m{p}}^2$
--- other choices will change the details of the graphs of
S$_{\m{AKV}}(\m{M_p^2\;;\;t}$) but not their basic shape.

We have made numerous trials with values of M$_{\m{p}}^2$ in
various ranges, and the results confirm our conjecture: for
example, Figure 4 shows the action function in a typical
``semi-exotic" case, M$_{\m{p}}^2$ = 1/3; we see that the action
does indeed become negative. Thus, as in the case where the $\psi$
field has a constant equation-of-state parameter, we cannot take
refuge in the semi-exotic regime: the only way to keep the brane
action positive is to go to the ``fully exotic" situation in which
the non-de Sitter part of the energy-momentum vector is itself
already spacelike.
\begin{figure}[!h]
\centering
\includegraphics[width=0.7\textwidth]{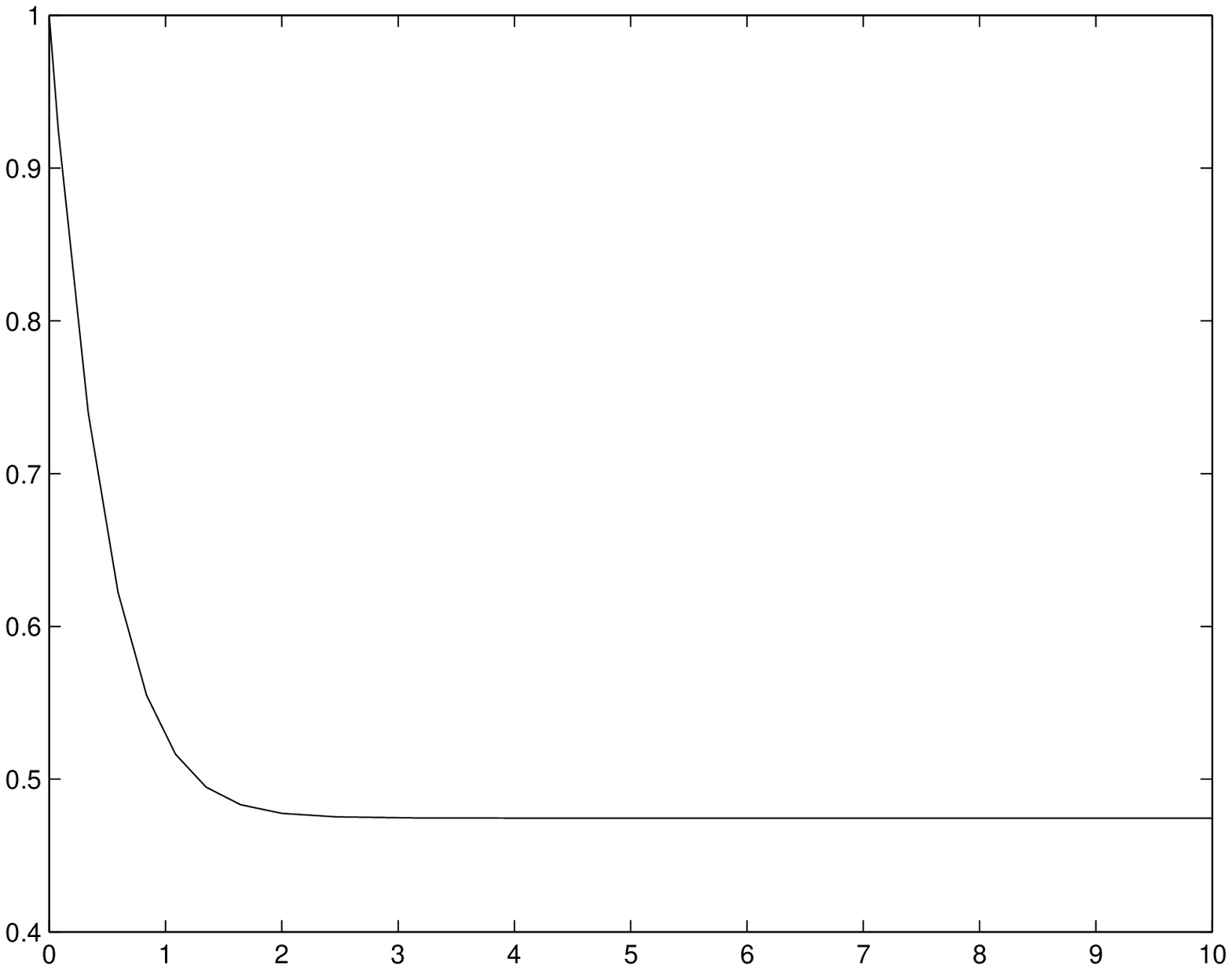}
\caption{The action S$_{\mathrm{AKV}}(1\;;\;\m{t})$.}
\end{figure}
Again, numerical trials show that this does work: see for example
Figure 5, which displays the action function when M$_{\m{p}}^2$ =
1. Thus the situation is the same whether or not we fix the
equation-of-state parameter of the non-de Sitter contribution to
be a constant. We believe that this conclusion holds for any
non-singular phantom cosmological model: brane pair-production
instability can only be avoided in the ``fully exotic" case,
corresponding to extremely fast decay of the non-de Sitter
component as the Universe expands.

\addtocounter{section}{1}
\section*{6. Conclusion}
We have three major conclusions.

First, string gas cosmologies must violate the NEC, at least
effectively, in their early histories, and this conclusion is very
robust: ultimately it is due to the \emph{topology} of the spatial
sections.

Second, this means that string gas cosmologies must be able to
cross the phantom divide, so models of inflation in string theory
must be compatible with such crossings; understanding this will no
doubt require a careful analysis of the internal degrees of
freedom in the dark energy generally. Of course, one has to ask
whether, in any case, an NEC-violating phase provides suitable
initial conditions for inflation; but this is not the place to
discuss that intricate question.

Thirdly, we saw that if a certain non-perturbative string effect
is taken into account, NEC violation in string gas cosmology can
only occur over brief periods: the NEC-violating ``field" [which
need not correspond to a true matter field] must dilute extremely
rapidly with the expansion of the Universe. This is just what we
want if we wish to account for w $<$ $-$1 in the extremely early
Universe, but it is not compatible with any such observation in
the relatively recent past. The phantom divide can be crossed only
once: the theory predicts that more precise observations of [for
example] supernovae will eventually indicate a dark energy
equation-of-state parameter w $\geq$ $-$1. See \cite{kn:tegmark}
for the prospects for this; note meanwhile that since it can be
argued that the current data \cite{kn:steinhardt} actually seem to
favour w $<$ $-$1, this is not a trivial statement.

The most striking aspect of this part of our discussion was that
\emph{it applies whether the NEC is violated in the true sense or
only effectively}. The particular kind of unstable production of
brane pairs which is relevant here is sensitive only to the
spacetime geometry --- the way in which that geometry is shaped
[whether by reversed-sign kinetic terms in a scalar field
lagrangian, or by brane-world models in which all matter fields
actually \emph{satisfy} the NEC] is quite irrelevant. Hence our
conclusion here --- that, according to string gas cosmology,
current indications of w $<$ $-$1 must be misleading --- is
robust.

The unusual features of string gas cosmology considered here stem
from the fact that these spacetimes are not singular. One
naturally wonders whether brane pair creation instability can be a
problem even in more conventional, initially singular cosmologies.
We shall investigate this question elsewhere.

We began this investigation with a discussion of the
Andersson-Galloway theorem \cite{kn:andergall}, which is clearly
the basic result on the spacetime structure of string gas
cosmologies. One of the basic assumptions of the theorem, however,
is that the spacetime in question is asymptotically locally de
Sitter. This is a reasonable assumption in the context of
inflationary cosmology, but it is not obvious that it is relevant
to the current phase of cosmic acceleration. It is entirely
possible that at some point in the relatively near future, the
current acceleration will cease, and be replaced by a contraction,
leading to a Big Crunch, as discussed in
\cite{kn:negative}\cite{kn:maoz}\cite{kn:mcinnes}\cite{kn:answering}\cite{kn:maoz3}\cite{kn:reallyflat};
see also \cite{kn:kratochvil} for the observational position. Note
that the models considered in
\cite{kn:mcinnes}\cite{kn:answering}\cite{kn:reallyflat} have flat
compact spatial sections, as in string gas cosmology. However, of
course, a string gas cosmology cannot have a Crunch, any more than
it can have a Bang. It would be interesting to understand the
details of the way in which string gas cosmologies avert a Crunch.

 \addtocounter{section}{1}
\section*{Acknowledgements}
The author is extremely grateful to Wanmei for the diagrams and
for unwavering support through difficult times. He would also like
to acknowledge the kind hospitality of the High Energy, Cosmology,
and Astroparticle Physics Section of the Abdus Salam International
Centre for Theoretical Physics, where much of the work described
here was done.

\end{document}